\documentstyle[12pt,epsfig]{article}

\begin {document}

\begin {center}
{\Large $\sigma$, $\kappa$ and $f_0(980)$ in E791 and BES\, II data}
\vskip 3mm
D.V. Bugg
\vskip 2mm
Queen Mary, University of London, Mile End Rd., London E1 4NS, UK
\end {center}

\begin{abstract}
Both $\sigma$ and $\kappa$ are well established from
E791 data on $D \to 3\pi$ and $K\pi \pi$ and BES II data on
$J/\Psi \to \omega \pi ^+\pi ^-$ and $K^+K^-\pi ^+\pi ^-$.
Fits to these data are accurately consistent with $\pi \pi$ and $K\pi$
elastic scattering when one allows for the Adler zero which arises from
Chiral Symmetry Breaking.
The phase variation with mass is also consistent between elastic
scattering and production data.
\end{abstract}


\vskip 4mm

At the conference, results on a large range of experiments on
the $\sigma$ and $\kappa$ were presented.
Here, available space limits the discussion  to E791 and
BES\, II data.
Space also prevents a review of the large background of
theoretical papers fitting elastic scattering.
An extended review covering all these topics will appear on
the web.

\vskip -4mm
%FIG 1.
\begin{figure} [htb]
\begin{center}
\centerline{\hspace{1.2cm}
\epsfig{file=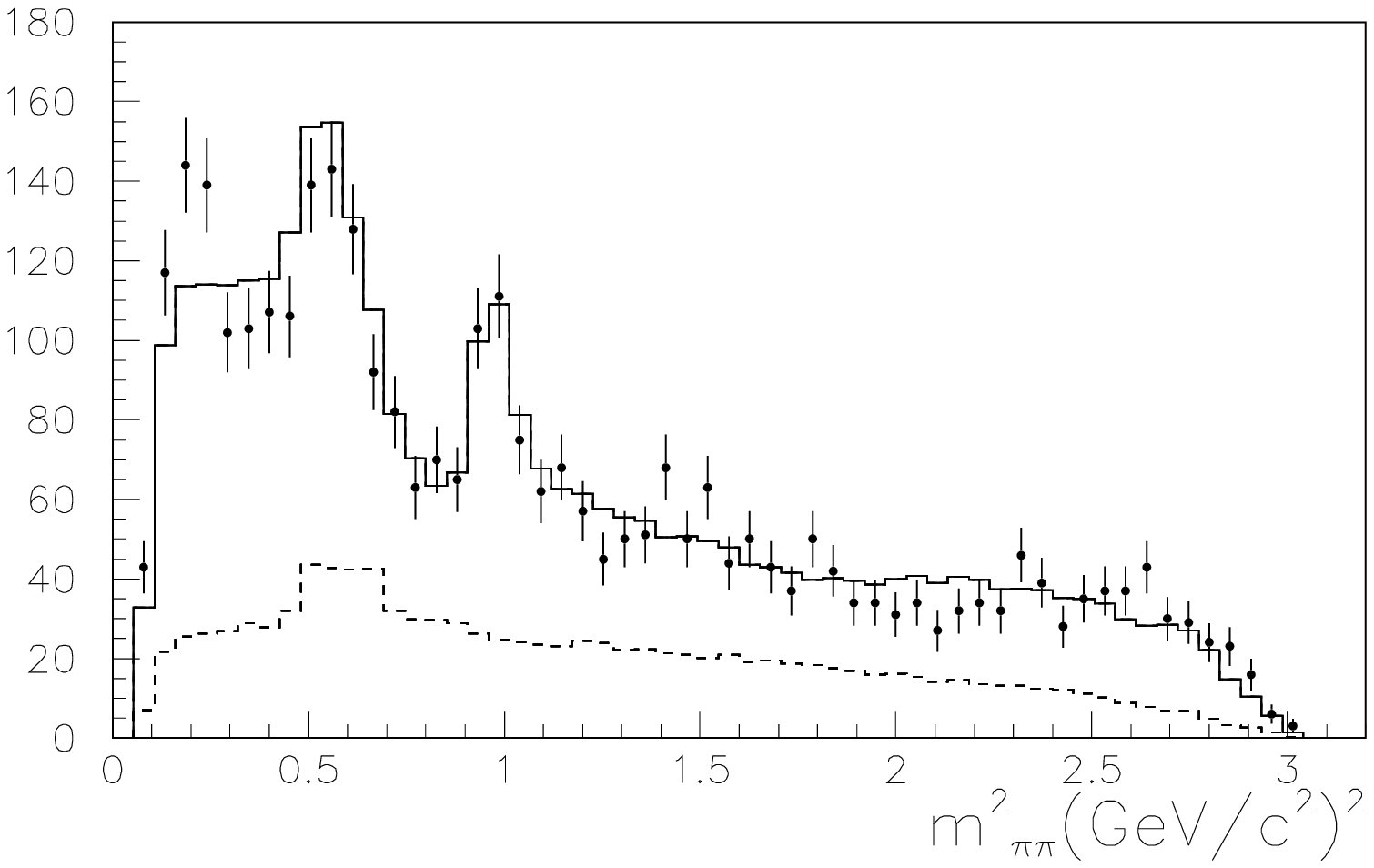,width=5.1cm}
\epsfig{file=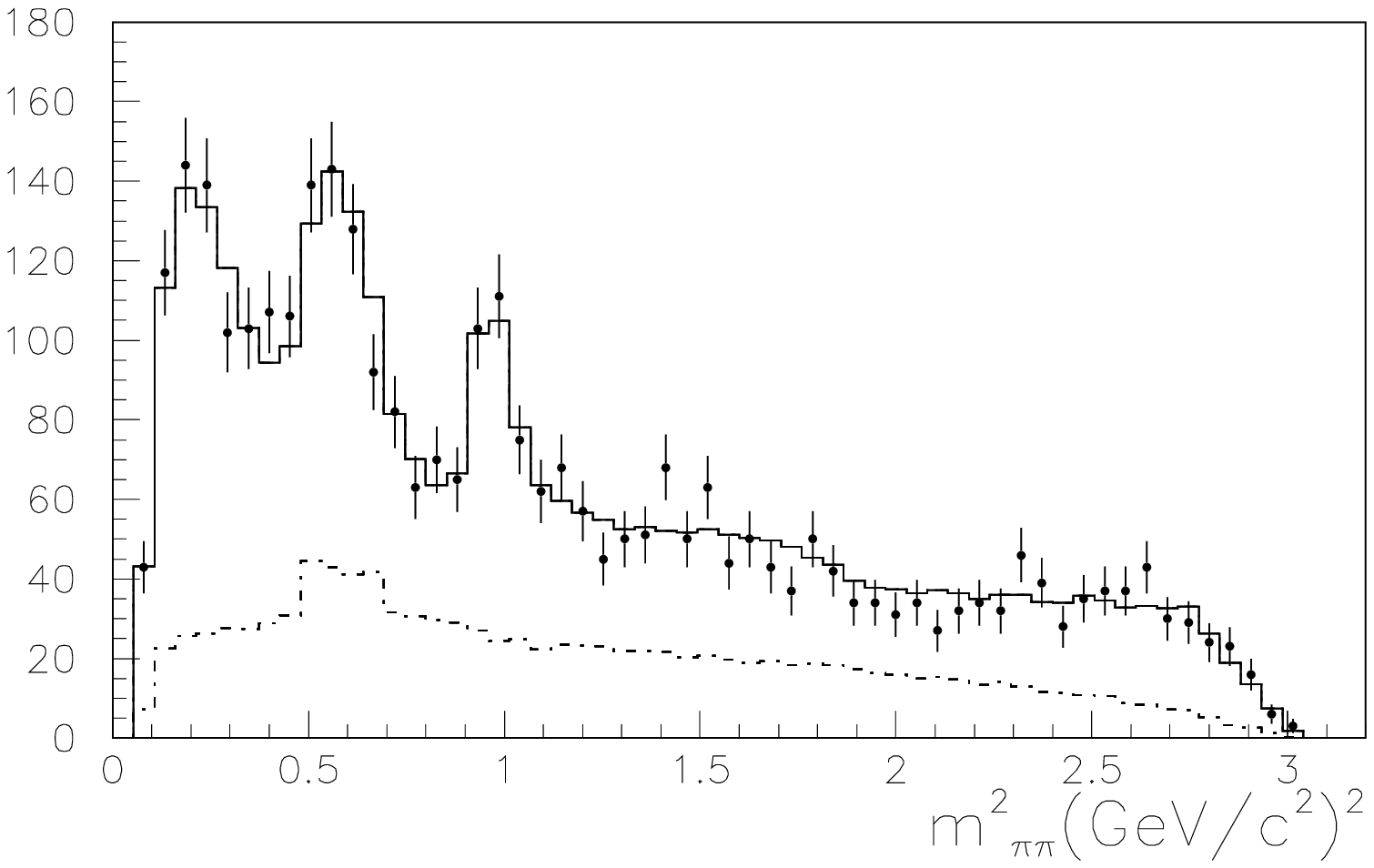,width=5.1cm}}
\vspace{-0.1cm}
\caption[]{ The $\pi \pi$ mass projection of E791 data for $D^+ \to \pi
^+(\pi ^- \pi ^+)$, (a) without, (b) with $\sigma$ in the fit.
The dashed histogram shows background.}
\end{center}
\end{figure}

\section {The $\sigma$ pole}
Early evidence for the $\sigma$ pole came from elastic scattering
data.
Markushin and Locher [1] summarise many determinations.
Renewed interest was sparked off by E791 data on $D^+ \to (\pi ^+ \pi
^-)\pi ^+$ [2].
The $\pi \pi$ mass projection, shown in Fig. 1, has a low mass peak.
The fit assumed a conventional Breit-Wigner resonance with
$\Gamma (s) \propto \rho (s)$, where $\rho (s)$ is Lorentz invariant
phase space $2k/\sqrt {s} = \sqrt {1 - 4m^2_\pi/s}$ and $k$ is
centre of mass momentum.
This choice of $\Gamma (s)$ will later be shown to be
inappropriate, but correcting it to a better form introduces only
changes of detail.
The pole position is shown in the first entry of Table 1 below.
Oller's very interesting refit of these data with ChPT input
is also shown there [3].

%FIG 2.
%\vskip -4 mm
\begin{figure} [t]
\begin{center}
\epsfig{file=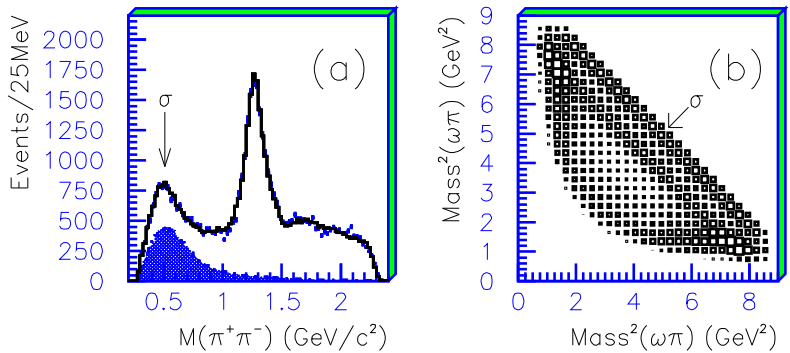,width=11.0cm}\
\vskip -6mm
\caption[]{ (a) The $\pi \pi$ mass projection of BES data;
the histogram shows the fit and the hatched area the $\sigma$
contribution; (b) the Dalitz plot. }
\end{center}
\end{figure}

Fig. 2 shows higher statistics BES\, II data [4] for $J/\Psi \to
\omega \pi ^+\pi ^-$.
Dominant signals are $f_2(1270)$, $b_1(1235)$ and $\sigma$, which
is clearly visible as a flat band along the right-hand edge of the
Dalitz plot in (b).
The $0^+$ contribution is shown shaded in (a).
All four parametrisations which were tried are consistent with an
average pole position $\rm {M} = (541 \pm 39) - i(252 \pm 42)$ MeV.

%\vskip -14mm
%FIG 3.
\begin{figure} [htb]
\begin{center}
\epsfig{file=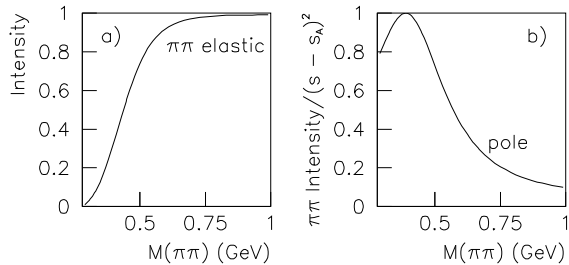,width=9.0cm}\
\vskip -8mm
\caption[]{ (a) The $\pi \pi$ intensity in elastic scattering,
(b) with the Adler zero divided out.}
\end{center}
\end{figure}

Fig. 3(a) shows the intensity of $\pi \pi$ elastic scattering v. mass.
The obvious puzzle is why there is no low mass peak like that in
production data of Fig. 2.
There is a simple explanation, given in 1965-6 by Adler and Weinberg
[5].
They proposed that massless $\pi$ of zero momentum have zero
scattering amplitude.
If the $I = 0$ S-wave $\pi \pi \to \pi \pi$ amplitude is expanded
as a power series $am^2_\pi +bk^2$, consistency between $s$, $t$ and
$u$ channels requires that the amplitude is proportional to
$(s - 0.5m^2_\pi)$ and has a zero at the Adler point
$s_A = 0.5 m^2_\pi$.
Fig. 3(b) shows the result of dividing (a) by $(s - s_A)^2$.
Instantly one sees a resemblance with the $\sigma$ peak of Fig. 2.
So the solution to the puzzle is that the matrix element for
$\pi \pi$ elastic scattering is strongly $s$-dependent: a situation
unlike most other resonances.

Let us write the elastic $\sigma$ amplitude as
%Eqn. 1
\begin {eqnarray}
\nonumber
T^{00}_{el} &=& [\eta \exp (2i\delta) - 1]/{2i} \\
            &=& \frac {N(s)}{D(s)} = \frac {N_{el}(s)}{M^2 - s -
            iN_{tot} (s)}.
\end {eqnarray}
Here $N_{el}(s)$ is real for $s \ge  0$; the phase variation comes
purely from the denominator $D(s)$.
This denominator is universal for all processes involving a $\pi\pi$
pair.
For elastic scattering, the Adler zero in $N(s)$ nearly cancels the
$\sigma$ pole for low masses.
However, the numerator $N(s)$ is not universal; it is quite
different for production processes, where the left-hand cut is distant.
Later, it will be shown that E791 data for $D \to (K\pi )\pi$ require
$N(s)_{prodn} = 1$ within errors.
The production amplitude will therefore be written
%Eqn. 2
\begin {equation}
T^{00}_{prodn}= \Lambda /D(s),
\end {equation}
where $\Lambda $ is a complex constant.

%\vskip -14mm
%FIG 4.
\begin{figure} [htb]
\begin{center}
\epsfig{file=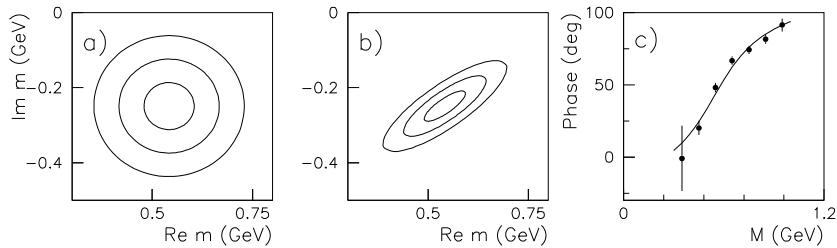,width=12cm}\
\vskip -4mm
\caption[]{ Contours of intensity for (a) production, (b) elastic
scattering; (c) the phase of the $\sigma$ amplitude in mass bins
100 MeV wide, compared with the global fit.}
\end{center}
\end{figure}

Fig. 4 illustrates contours of constant intensity for (a) production,
(b) elastic scattering.
In (b), the Adler zero suppresses the intensity near threshold.
The elastic phase shift on the real $s$ axis (where experiments are
done) reaches $90^\circ$ only at $M > 900$ MeV, far above the
pole. It is this feature which confuses many people.
The phase varies rapidly off the real axis because
the width of the resonance increases with $s$.

\begin{table} [t]
\begin{center}
\begin{tabular}{cccc}
State & Reference & Data & Pole position   \\
      &      &           & (MeV)            \\\hline
$\sigma$ & [2] & $D^+ \to (\pi ^+\pi ^- )\pi ^+$ & $(489 \pm 26) - i(173
            \pm 26)$ \\
      & [3] & $D^+ \to (\pi ^+\pi ^- )\pi ^+$ & $470 - i220$   \\
      & [4] &  $J/\Psi \to \omega (\pi ^+\pi ^- )$ &
           $(541 \pm 39) - i(252 \pm 42)$   \\
      & [10] & $\pi \pi \to \pi \pi$ & $(470 \pm 30) - i(295 \pm 20)$
      \\
      & [13] & $\pi \pi \to \pi \pi$ & $445 - i221$ \\
      & [25] & $\pi \pi \to \pi \pi$ & $(470 \pm 50) - i(285 \pm 50)$ \\
$\kappa$ & [28] & $D^+ \to (K^-\pi ^+)\pi ^+$
         & $(721 \pm 61) - i(292 \pm 131)$\\
         & [30] & $J/\Psi \to K^+\pi ^- K^-\pi ^+$ & $(760 \pm 41) -
         i(420 \pm 75)$ \\
         & [29] & $J/\Psi \to K^+\pi ^- K^-\pi ^+$ & $(841 \pm 82) -
         i(309 \pm 87)$ \\
         & [33] & $K\pi \to K\pi$ & $(722 \pm 60) -
         i(386 \pm 50)$ \\
         & here & all & $750^{+30}_{-55} -i(342 \pm 60)$ \\
         & [3]  &$D^+ \to (K^-\pi ^+)\pi ^+$ & $710 -i310$ \\
         & [12] & $K\pi \to K\pi$ & $770 - i(250-425)$ \\
         & [14] & $K\pi \to K\pi$ & $708 - i305$ \\
         & [15] & $K\pi \to K\pi$ & $753 - i235$ \\
         & [24] & $K\pi \to K\pi$ & $(594 \pm 79) - i(362 \pm 322)$ \\
$f_0(980)$& [34] & $J/\Psi \to \phi \pi ^+ \pi ^-$ & $(998 \pm 4) - i(
           17 \pm 4)$ \\
         & [12] & $\pi \pi \to \pi \pi$ and $KK$ & 994 - i14 \\
$a_0(980)$& [35] & $\bar pp \to \eta \pi \pi$ and $\omega \eta \pi ^0$
         & $(1036 \pm 5) - i(84 \pm 9)$ \\\hline
\end{tabular}
\caption {Summary of pole positions.}
\end{center}
\end{table}

Data from BES\, II, $K_{e4}$ [6] and
Cern-Munich [7]   are fitted empirically with:
%Eqn. 3
\begin {equation}
N(s) = M(s - 0.5m^2_\pi )\exp[-(s - M^2)/A](1 + \beta s) \rho _{\pi
\pi}(s) + M\Gamma _{4\pi }(s).
\end {equation}
The exponential is required by $\pi \pi$ elastic data to cut off $N(s)$
above 1 GeV.
The term $M\Gamma _{4\pi }$ fits inelasticity above 1 GeV, but
has little effect on the $\sigma$ pole.
Note from Fig. 2(a) that the $\sigma$ intensity fitted to BES data
is small above 1 GeV.
The $\sigma$ pole is therefore distinct from the
broad $f_0(1535)$ fitted by Anisovich and Sarantsev [8].
It is also  distinct from the broad pole fitted around 1 GeV by
Au, Morgan and Pennington [9].
Contributions from $f_0(980)$, $f_0(1370)$ and $f_0(1500)$ to elastic
scattering are included by multiplying their S-matrices with
$S_{\sigma }$, to satisfy unitarity; i.e. their {\it phases} add.
For production, there are hundreds of open channels for
$D$ and $J/\Psi$ decays. Within individual channels, unitarity plays a
negligible role.
Following the standard isobar model, {\it amplitudes}
are added using a complex coupling constant $\Lambda = g\exp (i\phi
_0)$ for each amplitude.

The $K_{e4}$ data are available up to 380 MeV and there is then a
gap in elastic data until 560 MeV, where Cern-Munich data begin.
The $\sigma$ pole lies in the mass range where there are no
elastic data.
Although this gap may be bridged by using dispersion relations,
the production data are obviously important in filling the gap directly.

%\subsection {The phase of the $\sigma$}
In the BES data, $b_1\pi$ contributes 41\% of the
intensity and $\sigma \omega$ 19\%.
Strong interferences between them determine the
phase variation of the $\sigma$ with mass.
The data have been divided into bins 100 MeV wide from 400 to 1000 MeV.
Unfortunately, the $b_1$ band runs off the corner of the Dalitz
plot and does not interfere significantly with the $\sigma$ below
400 MeV.
Fig. 4(c) shows phases for individual bins, in good
agreement with the global fit.

\subsection {Theory}
Colangelo, Gasser and Leutwyler [10] have made a precise determination
of the $\sigma$ pole from elastic data and $K_{e4}$
without using production data.
This work has the virtue of fitting both the physical region and
the left-hand cut.

Oset, Oller, Pelaez and collaborators fit  elastic
data using `unitarised' Chiral Perturbation Theory [11-16].
They use ChPT for lowest order and take rescattering from the next
order.
They fit successfully not only the $I = 0$ S-wave, but
also the repulsive $I = 2$ S-wave.
This implies that they fit the nearby left-hand cut correctly as well
as the physical region.

Schechter's group has also contributed a series of papers on
all of $\sigma$, $\kappa$, $f_0(980)$ and $a_0(980)$ [17--21]. This
work examines possible mixing between 2-quark and 4-quark states.
Zheng and collaborators at Peking University have developed new
types of dispersion relations and have applied them to analysis
of data on the $\sigma$ and $\kappa$ [22--25].

Van Beveren and Rupp have an interesting but quite different scheme,
modelling the spectrum and decays of all mesons from the lightest to
charmonium and bottomonium states. This is done with a harmonic
oscillator [26] or an arbitrary confining potential [27].
They allow for decays by coupling q-qbar states to outgoing mesons
through a transition potential.
In this scheme, $f_0(1370)$, $a_0(1450)$, etc. are regular but
unitarised $q\bar q$ states;
$\sigma$, $\kappa$, $f_0(980)$ and $a_0(980)$ appear as `extra'
states created by coupling of $q\bar q$ to decay channels [26,27].

%FIG 5.
\begin{figure}
\begin{center}
\centerline{\hspace{1.2cm}
\epsfig{file=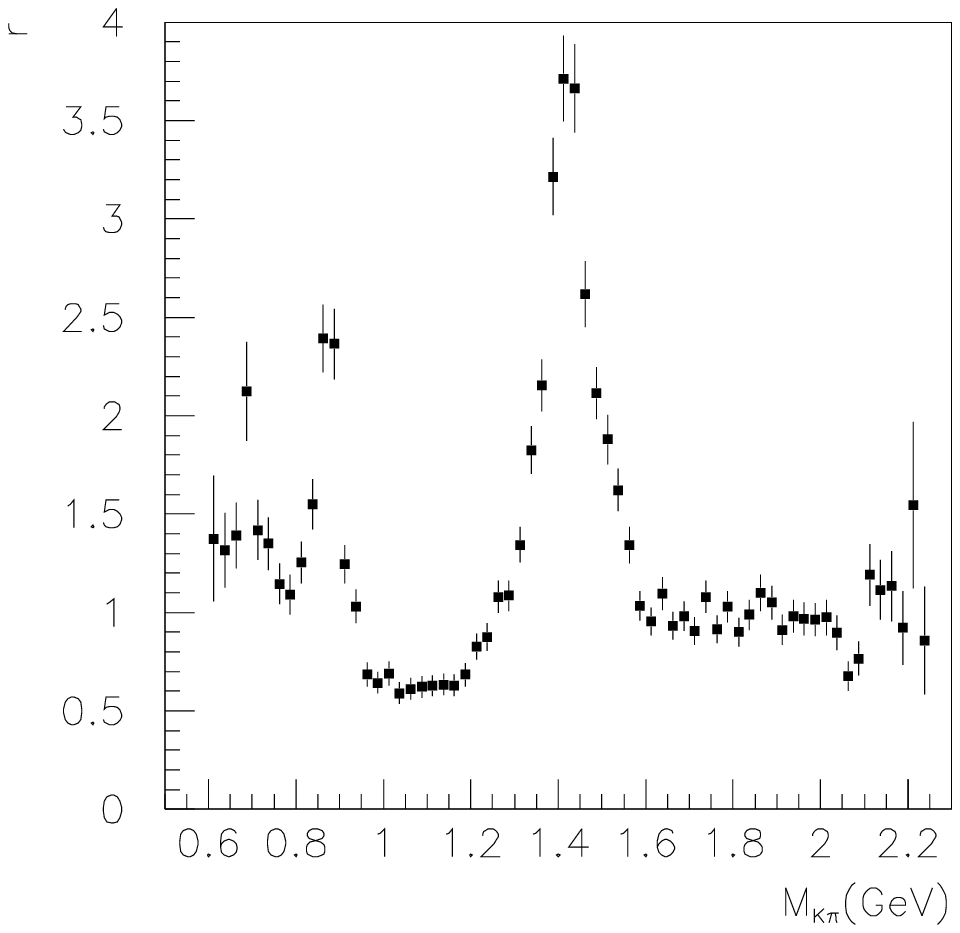,width=4.1cm}
\epsfig{file=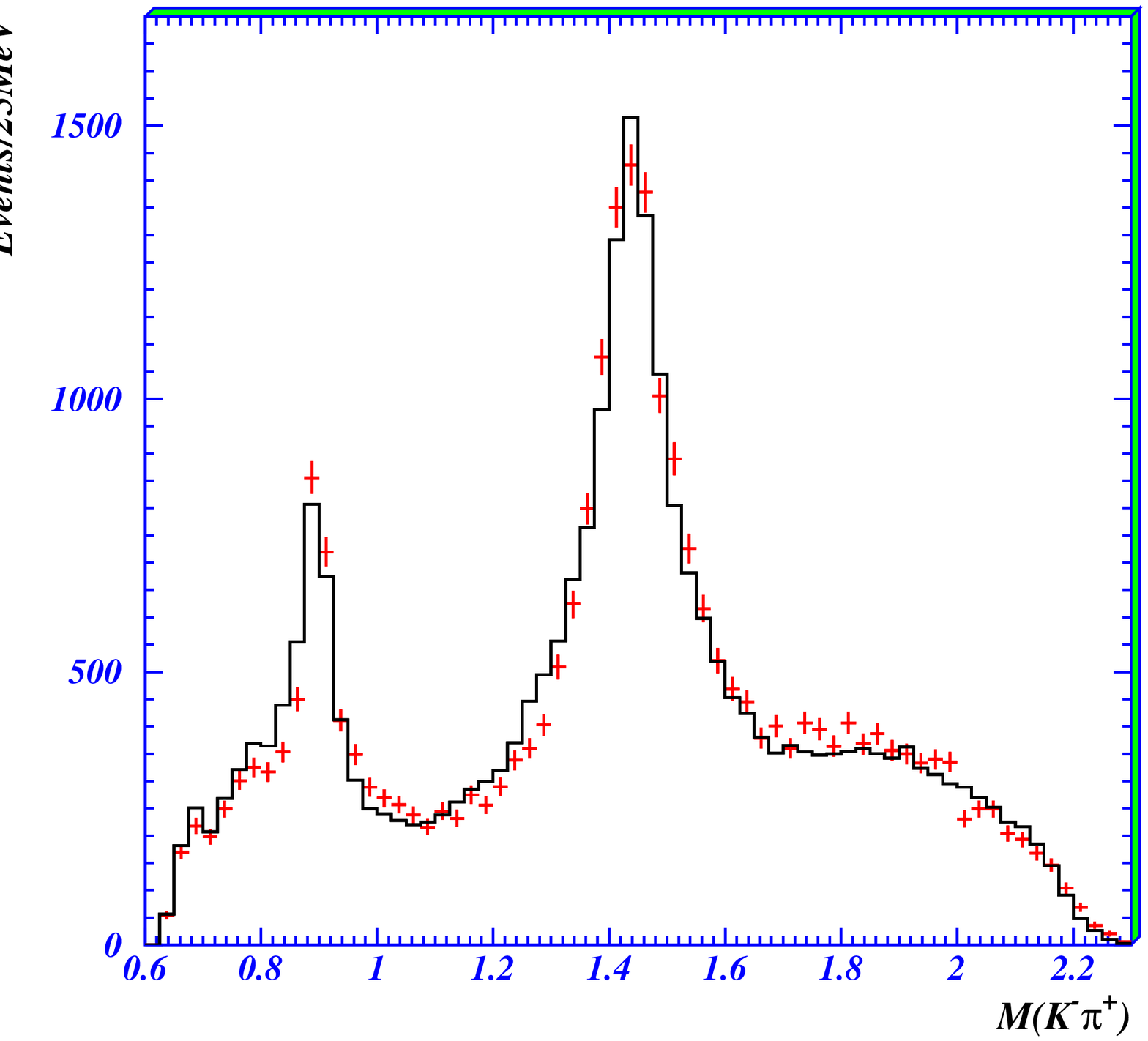,width=5.1cm}}
\vspace{0.1cm}
\centerline{\hspace{0.3cm}
\epsfig{file=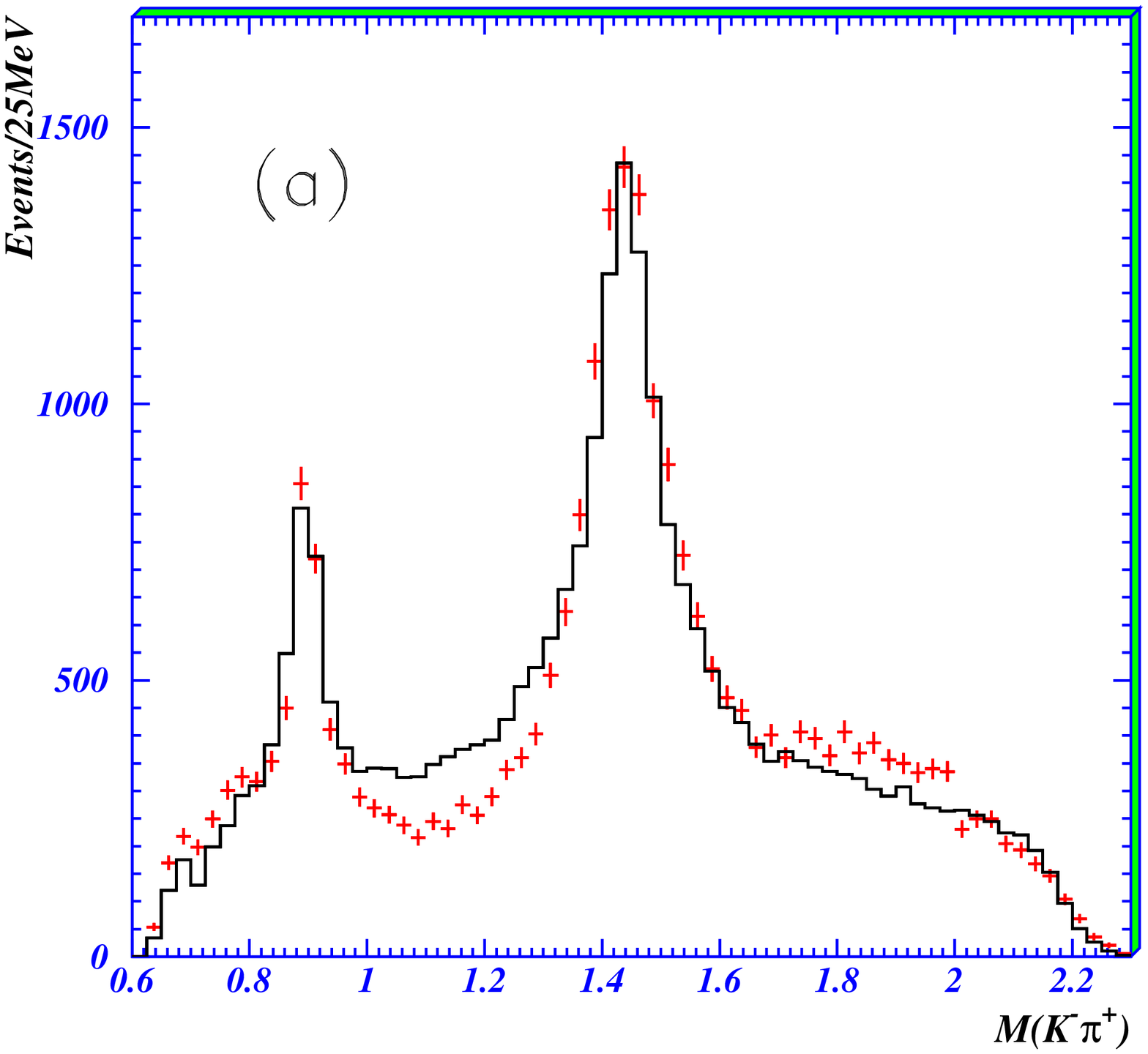,width=5.1cm}
\epsfig{file=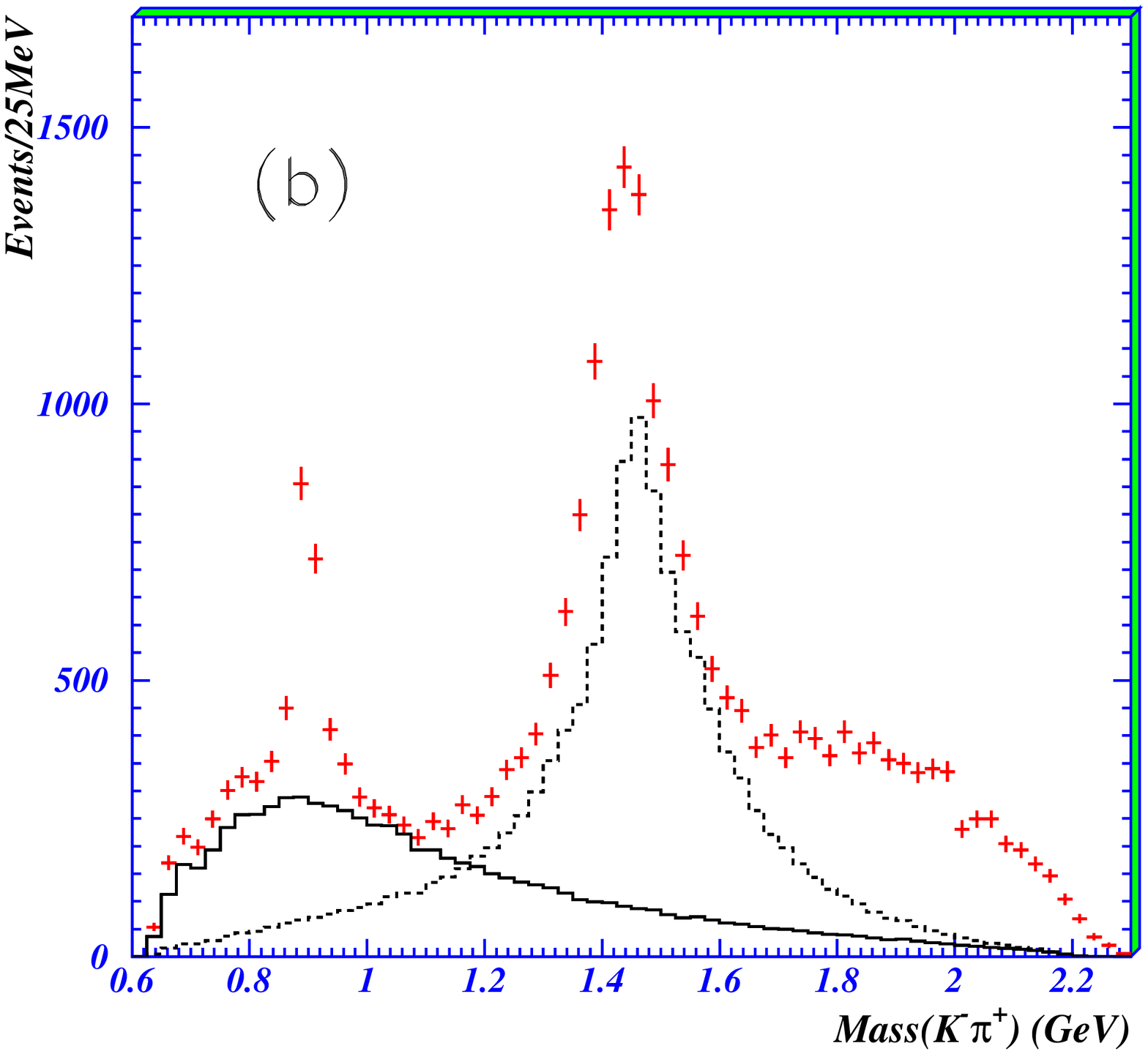,width=5.1cm}}
\vspace{-0.1cm}

\caption[]{ BES\, II data on the $\kappa$.
Upper left: the $K\pi$ S-wave spectrum in $J/\Psi \to K^*(890)(K\pi )_S$
after factoring out phase space; upper right:  the fit to $m(K\pi)$;
(a) the bad fit if the Adler zero is omitted; (b) the
$\kappa$ contribution (full histogram) and $K^0(1430)$ (dashed).}
\end{center}
\end{figure}

\section {The $\kappa$ pole}
E791 data on $D^+ \to (K^-\pi ^+)\pi ^+$ provided the first evidence for
the $\kappa$ pole from production data [28].
A combined fit to these and other data will be considered in
detail below.

Next, BES\, II data on $J/\Psi \to K^+\pi ^- K^-\pi ^+$
reveal the $\kappa$ in $J/\Psi \to K^*(890)\kappa $ [29,30].
If one selects a $K^\pm \pi ^\mp$ pair within 50 MeV of 890 MeV,
the accompanying $K^\mp \pi ^\pm$ pair has the mass projection shown
in Fig. 5(a) after dividing out phase space.
The broad low mass peak below the $K^*(890)$ is the evidence for
the $\kappa$; one point has a rather large statistical fluctuation.
Full details of my fit to these data are given in Ref. [30].
Other strong components are
$K^*(890)K_0(1430)$, $K^*(890)K_2(1430)$, $K_0(1430)$$K_0(1430)$
and $KK_1(1270)+KK_1(1400)$, followed by $K_1 \to K^*\pi$ and $\rho K$.
Since $K_1$ decays populate the low mass $K\pi$ range, it is
essential to demonstrate that $K_1$ decays do not reproduce the
$\kappa$ peak.

%\vskip -10mm
%FIG 6.
\begin{figure} [htb]
\begin{center}
\epsfig{file=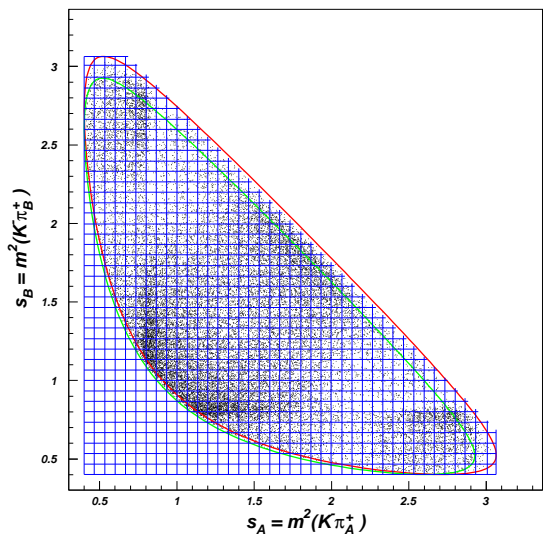,width=8.0cm}\
\vskip -5mm
\caption[]{ The Dalitz plot for E791 data for $D^+ \to K^-\pi ^+\pi
^+$.}
\end{center}
\end{figure}

A combined fit is made to BES and LASS data [31] for the
$K\pi$ $I = 1/2$ S-wave.
The fit uses eqn. (1) with
%Eqn. (8)
\begin {equation}
N(s) = M(s - s_A)\exp (-\alpha \sqrt {s})\rho_{K\pi }(s)
\end {equation}
and $s_A = m^2_K - 0.5 m^2_\pi$.
The fit to the raw $K\pi$ mass projection is shown in Fig. 5(b).
If the factor $(s - s_A)$ is omitted, the poor fit is shown in
Fig. 5(c).
The $\kappa$ and $K_0(1430)$ mass projections are shown in Fig.
5(d).
There is destructive interference between them.
Sensitivity to this interference is one reason for fitting
the LASS data simultaneously.
A second point is to show that
eqn. (4) fits both sets of data successfully.

The phase variation of the $\kappa$ with mass is well determined
in two ways which agree.
Firstly, there is large interference between channels
$K^*(980)\kappa$ and $KK_1(1270+1400)$.
Secondly, there is a large interference between $\kappa$ and
$K_0(1430)$, which both contribute to the $K\pi$ S-wave.
A bin-by-bin fit has been made where the $\kappa$ signal is
fitted in magnitude and phase in 10 individual bins 100 MeV wide.
Results are shown in Fig. 8(c) below.

An independent analysis of exactly the same data has been reported by
the BES group [29].
This fit omits interferences of the  $\kappa$ with both $K_0(1430)$
and the $K_1$'s. It therefore provides no information on the
phase of the $\kappa$.
It also does not fit $K^*(890)$ decays; angular correlations with
these decays are crucial in separating channels $K^*(890)K_0(1430)$ and
$K^*(890)K_2(1430)$.

\section {Re-analysis of E791 data}
The Dalitz plot for E791 data in shown in Fig. 6.
There is obvious interference between $K^*(890)$ and the
$K\pi$ S-wave, creating an asymmetry around the $K^*$ band.
A  new fit has been reported recently where the magnitude and
phase of the $K\pi$ S-wave anplitude is fitted separately in 37 mass
bins [32].
Magnitudes and phases are shown below in Figs. 7 and 8.
No attempt is made to separate $\kappa$ and $K_0(1430)$.

Both E791 and LASS data  contain  fairly weak $K_0(1430)$
structures. Results depend the variation of $\kappa$ intensity
under the $K_0(1430)$; i.e. the width of the $\kappa$ is
correlated with the fit to $K_0(1430)$. I have made a combined fit to
the data of  LASS, BES and E791. The BES data define well the
$K_0(1430)$ peak, which is much more conspicuous than in either LASS or
E791 data.

%\vskip -8mm
%FIG 7.
\begin{figure} [htb]
\begin{center}
\epsfig{file=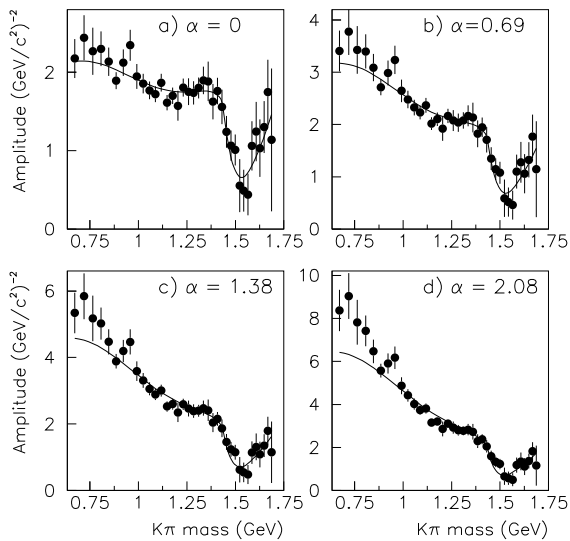,width=8.0cm}\
\vskip -3mm
\caption[]{ The fit to the magnitude of the $K\pi$ S-wave in E791
data for four values of $\alpha$ in the production form factor.}
\end{center}
\end{figure}

In the E791 fit (and also
Ref. [28]), the amplitude includes a production form factor
$F = \exp (-\alpha q^2)$, where $q$ is the momentum of
the $\kappa$ in the $D$ rest frame and $\alpha = 2.08$ GeV$^{-2}$.
I have varied  $\alpha$ and Fig. 7 shows results.
Panel (a) uses $\alpha = 0$ and gives the best fit.
In (b)--(d) $\alpha$ increases in equal steps to the E791 value
in (d). Within errors, $\alpha$ optimises at 0.
This corresponds to a point-like decay $D \to
\kappa \pi$ with an RMS radius $< 0.38$ fm with 95\% confidence.
Fits to the phase of the $\kappa$ also optimise at
$\alpha = 0$.
The combined fit gives a $\kappa$ pole at
$\rm {M} = (750 ^{+30}_{-55}) - i(342 \pm 60)$ MeV.
This compares with $\rm {M} = (722 \pm 60) -i(386 \pm 50)$ MeV for
LASS data alone [33]
and ${\rm M} = (760 \pm 20\pm 40)
         -i(420 \pm 45 \pm 60)$ MeV for LASS + BES data only [30].
It reduces errors and brings all three sets of data into close
agreement.

%\vskip -6mm
%FIG 8.
\begin{figure} [htb]
\begin{center}
\epsfig{file=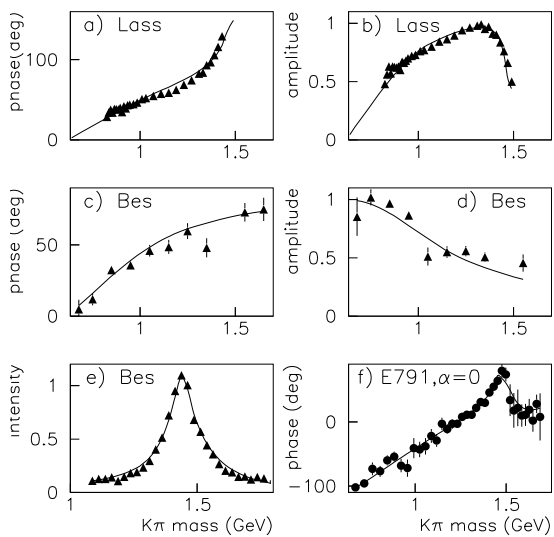,width=8.0cm}\
\vskip -3mm
\caption[]{ Fits to (a) Lass phases for $K\pi$ elastic scattering;
(b) magnitudes of the Lass elastic scattering amplitude;
(c) the phase of the $\kappa$ signal in BES\, II data after
dividing into 100 MeV mass bins, (d) BES magnitudes in individual bins,
(e) the magnitude of the $K_0(1430)$ peak above background in BES
data,
(f) E791 phases with $\alpha = 0$.}
\end{center}
\end{figure}

Fits to the BES  and LASS data are shown in Fig. 8.
Figs. 8(a) and (b) show the fit to LASS data;
(c) and (d) show that the
fit to the magnitude and phase of the $\kappa$ in BES data agrees.
Again, the phase variation with mass agrees between
elastic scattering and production data.
This demonstrates that both can be fitted with the same
$D(s)$, and therefore that the data can be fitted with
a resonance alone without the need for any background amplitude.
The $K_0(1430)$ is a large signal in BES data with well defined
centroid and width.
The fit to this peak is shown in Fig. 8(e) using
25\% $K^*(890)K_2(1430)$ and 75\% $K^*(890)K_0(1430)$,
as determined from my fit to BES data;
(f) shows the good fit to E791 phases using $\alpha = 0$.

\section {$f_0(980)$ and $a_0(980)$}
An excellent new determination of the parameters of $f_0(980)$
is obtained from BES data on $J/\Psi \to \phi \pi ^+\pi ^-$
and $\phi K^+K^-$ [34].
The $f_0(980)$ appears as a strong peak in
$\pi \pi$ and a smaller but clear peak in $KK$.
From the relative numbers of events in these two peaks, the ratio
$g^2(KK)/g^2(\pi \pi )$ is well determined.
Previously results for this ratio have been erratic.
The pole position is shown in Table 1, together with the best
available pole position for $a_0(980)$ [35].

\section {Interpretation of $\sigma$, $\kappa$, $f_0(980)$ and
$a_0(980)$}
This is contentious.
The front runner appears to be Jaffe's proposal [36] that they are
4-quark states with $SU(3)$ flavour content $3 \otimes \bar 3$.
His scheme explains naturally the progression of their masses:
if $f_0(980)$ and $a_0(980)$ are $s\bar s (u\bar u \pm d\bar d)$
states, their near degeneracy is explained naturally, and the
large mass difference between $\sigma$ and $a_0(980)$.
There is support for this interpretation from Lattice QCD calculations,
where 4-quark combinations dominate at large $r$ and 2-quark
combinations at small $r$ [37]. The decay presumably occurs by
fission at small $r$. There is again support for this picture from
Ref. [33], where it is shown that the Fourier transform of the
$\sigma$ amplitude v, momentum has an RMS radius of 0.4 fm.

However, the story cannot be quite so simple.
The ratio $r = g^2(f_0(980) \to KK)/g^2(a_0(980) \to KK) = 2.7 \pm 0.5$
disagrees with the ratio 1 predicted from Jaffe's model.
The likely explanation is that the pole position of $f_0(980)$
is fortuitously very close to the $KK$ threshold, and that of $a_0(980)$
lies further away.
The $f_0(980)$ then necessarily has a large long-range $KK$ cloud,
larger than $a_0(980)$.
Decays to $\pi \pi$ or $\pi \eta$ occur at small $r$ (and some $KK$);
when this happens, the $KK$ cloud at large $r$ is left `in the air'
(adiabatic approximation).
This results in fall-apart decay which is larger for $f_0(980)$
than $a_0(980)$.
This argument accounts for a factor 2 increase in $KK$ decays of
$f_0(980)$ compared to $a_0(980)$,
but not the precise pole positions of $f_0(980)$ and $a_0(980)$.

\end {document}